\documentclass[aps,twocolumn]{revtex4}  
\usepackage{graphicx}

\usepackage{amsfonts, amsmath, amssymb} 

\newcommand{\h}{\overline{\text{H}}}

\begin{document}

\title[Lifetime of Trapped Antihydrogen]{Lifetime of Magnetically Trapped Antihydrogen in ALPHA}

\author{A. Capra}
\email{acapra@triumf.ca}
\affiliation{TRIUMF, 4004 Wesbrook Mall, Vancouver, BC, Canada V6T 2A3}
\author{\mbox{on behalf of the} ALPHA collaboration}

\date{18 June 2018}

\begin{abstract}
How long antihydrogen atoms linger in the ALPHA magnetic trap is an important characteristic of the ALPHA apparatus. The initial trapping experiments in 2010 \cite{AndreNat2010} were conducted with 38 detected antiatoms confined for 172 ms and in 2011 \cite{AndreNat2011} with seven for 1000 s. Long confinement times are necessary to perform detailed frequency scans during spectroscopic measurements. An analysis carried out, using machine learning methods, on more than 1000 antiatoms confined for several hours in the ALPHA-2 magnetic trap, yields a preliminary lower limit to the lifetime of 66 hours. Hence this observation suggests that the measured confinement time of antihydrogen is extended by more than two orders of magnitude.
\end{abstract}

\maketitle

\section{Introduction}
\label{intro}
\begin{figure}[!h]
 \includegraphics[scale=0.12]{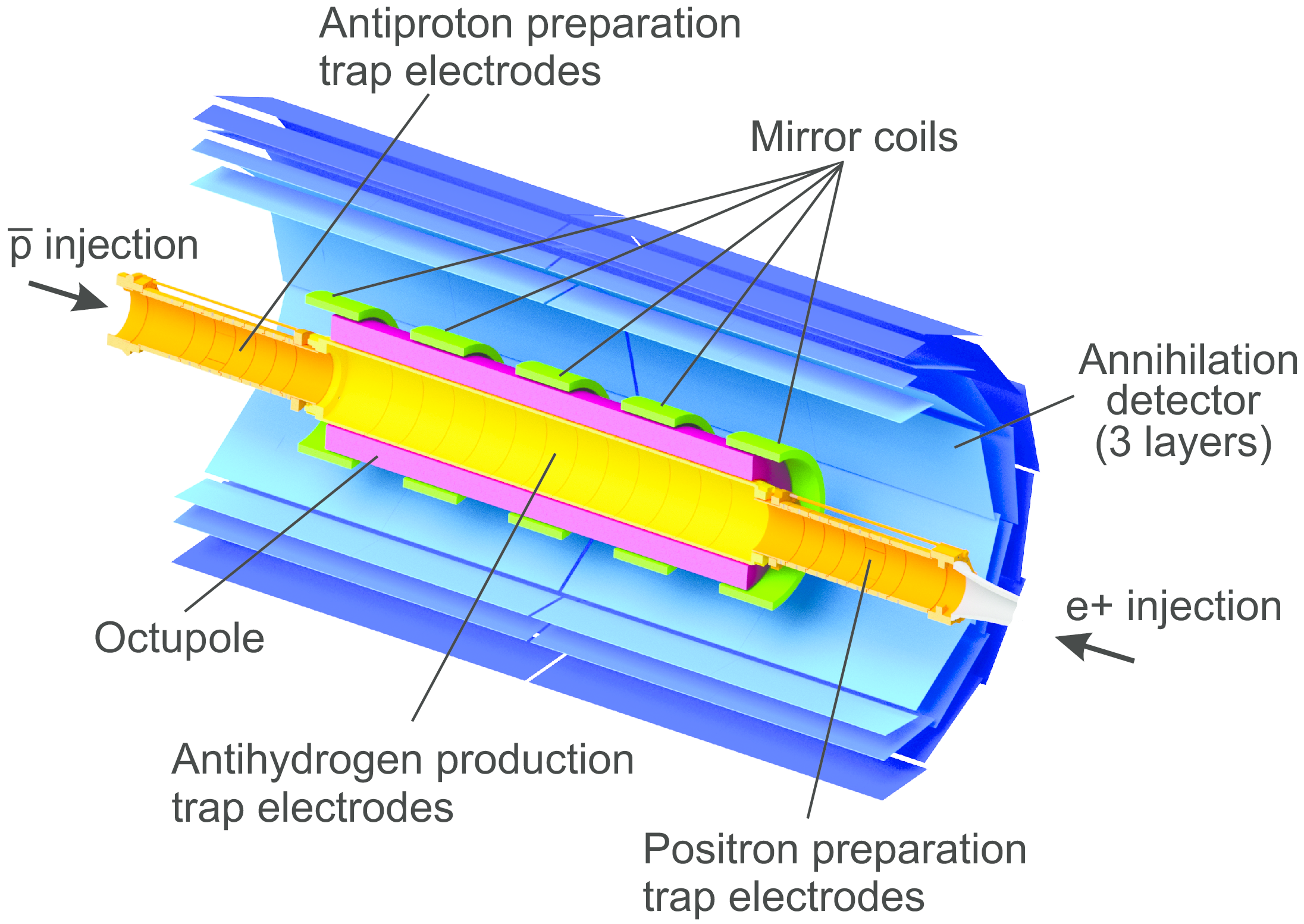}
 \caption{Sketch of the ALPHA apparatus, showing the main components.}
\label{fig:ALPHAapp}       
\end{figure}

The ALPHA apparatus at the CERN Antiproton Decelerator (AD) produces, traps and studies cold antihydrogen atoms, or $\h$ (see Fig.~\ref{fig:ALPHAapp}). The $\h$ synthesis is accomplished by mixing cold plasmas of antitprotons and positrons in a Malmberg-Penning trap. The $\h$ confinement is achieved by exploiting its magnetic moment and employing a set of superconducting coils: the so-called \textit{mirror coils} provide the axial confinement, while an \textit{octupole} magnet is used to trap the antiatoms radially. These magnets produce a magnetic field gradient and the low-field seekers $\h$ are confined near its minimum. $\h$ is studied by exciting the 1S-2S transition \cite{a17obs1s2s}\cite{a18char1s2s} and the hyperfine Zeeman levels of the ground state \cite{a17obshf}. These studies aim to test the Charge-Parity-Time symmetry by comparing the transitions frequencies of $\h$ to the well-known ones in hydrogen.

One of key features of the ALPHA apparatus is the \textit{silicon vertex detector}, or SVD, which is employed to identify $\h$ annihilation by reconstructing the trajectories of the charged particles produced in the antiproton annihilation.

The single-particle sensitivity of the SVD is crucial to probe the interaction of $\h$ with the electromagnetic radiation during the aforementioned spectroscopic experiments. During the exposure to the laser light or the microwave radiation, the antitoms are held in the magnetic trap. When, for instance, the antiatoms interact with resonant radiation, the number of $\h$ annihilation detected by the SVD during this phase is non-zero. Typical illumination times range from few minutes to few hours, after which the magnets are ramped down. The remaining $\h$ are ejected radially and counted in the SVD.

In recent years, the average number of antiatoms confined in each experimental cycle has increased dramatically thanks to  better control of the temperature and size of the non-neutral plasmas \cite{a18enh}, used for $\h$ production, and to the development of a technique to synthesize and accumulate antiatoms \cite{a17acc} over multiple spills of the AD, which delivers $\approx3\times10^{7}$ antiprotons every $\approx100\,$s. It is easily conceivable that a large number of trapped $\h$ lead to better statics for any measurement, such as the one presented here.

\section{Description of the Lifetime Experiment}
\begin{figure}[!h]
\includegraphics[scale=0.1]{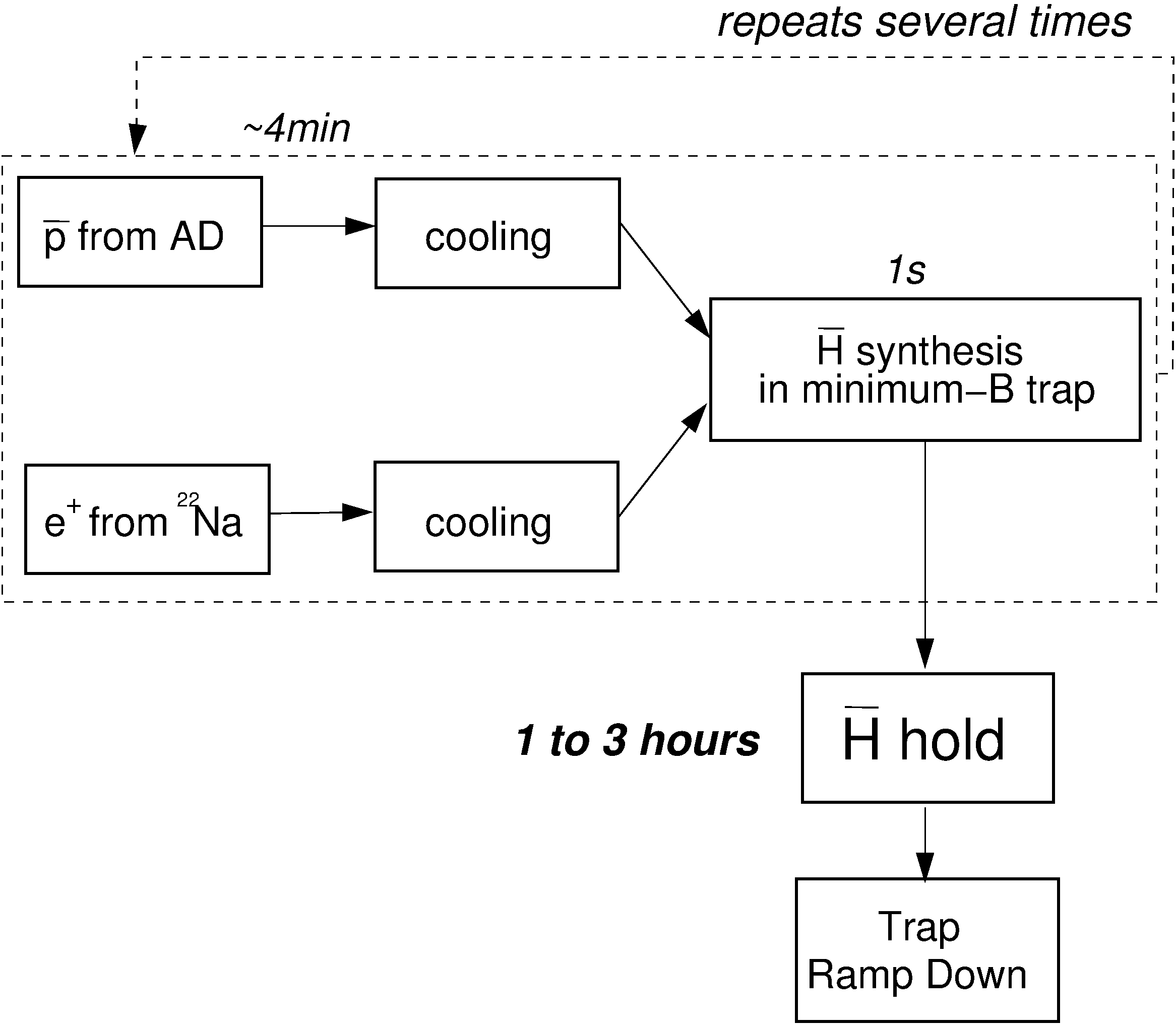}
 \caption{Simplified flow chart of the experimental procedure.}
\label{fig:Exptcyc}       
\end{figure}

Four datasets have been collected (see Table~\ref{tab:exppar}) following the experimental procedure shown in Fig.~\ref{fig:Exptcyc}. 

After $\h$ is produced in the neutral (anti)atom trap, by mixing cold plasmas of antiprotons and positrons, it is confined near the minimum of the inhomogeneous magnetic field generated by superconducting magnets. The $\h$ synthesis, or \textit{mixing}, is repeated several time to accumulate a large number of antiatoms. These antitoms are held in the magnetic trap for few hours, for spectroscopy purpose, during the so-called \textit{hold phase}. In the present measurement, no laser light was present in the $\h$ confinement region. The trap is then turned off by ramping down the superconducting magnets, called \textit{FRD phase}, over few seconds, hence releasing $\h$ over the same period of time.

\begin{table}[!h]
\caption{The four datasets differ by the number of (successful) $\h$ synthesis cycles, by the duration of the hold phase and by the duration of the FRD - or \textit{ramp down} - phase.}
\label{tab:exppar}       
\resizebox{0.5\textwidth}{!}{  
\begin{tabular}{lllll}
\hline\noalign{\smallskip}
\multicolumn{5}{c}{\textbf{Experimental Parameters}} \\
\hline\noalign{\smallskip}
                                                                  &\textbf{\#1} &\textbf{\#2} &\textbf{\#3} &\textbf{\#4}\\
\noalign{\smallskip}\hline\noalign{\smallskip}
Number of mixing cycles                                             &29        &11        &23        &46\\
$T$ \textit{hold time} in s                                         &7530      &11230     &3770      &3770\\
Expected background events \textit{hold phase} $R\times T$	    &34        &51        &17        &17\\
Selected events in \textit{hold phase}                              &35        &53        &14        &19\\
Number of events in \textit{ramp down} phase                        &365       &177       &145       &224\\
Expected background $B_{\text{FRD}}$ in \textit{ramp down} phase    &0.098     &0.498     &0.733     &0.733\\
\noalign{\smallskip}\hline
\end{tabular}
}
\end{table}

During the \textit{hold phase}, the loss of trapped $\h$ is monitored by the SVD. The sensitivity to the loss rate is due to the comparison between the number of events selected during the hold phase and the ones selected when the trap is shut down (and emptied).

\section{Antihydrogen Detection and Background Rejection}
\label{sec:mva}
The main source of background to $\h$ annihilation detection is due to cosmic rays, which are triggered at a rate of $10\,$Hz. In order to perform experiments during an extended period of time and to boost their statistical significance, the background must be reduced by several orders of magnitude. The use of Machine Learning (ML) algorithms has allowed ALPHA to cut down the cosmic ray rate by more than three orders of magnitude. Other sources of background are eliminated by employing other techniques, not discussed here.

\begin{figure}[!h]
\includegraphics[scale=0.12]{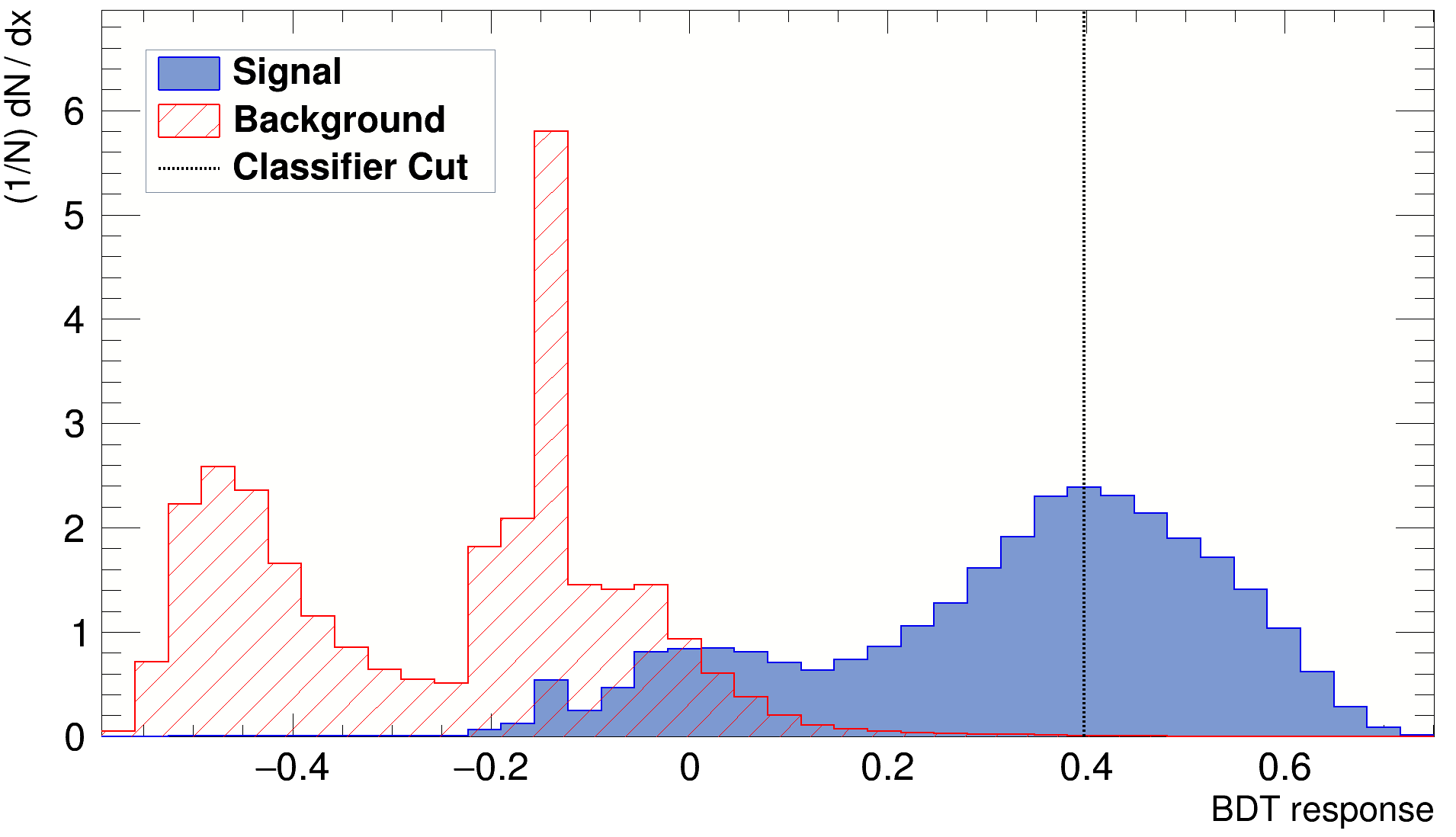}
\caption{Response of the chosen classifier on the test sample. The dashed line indicates the cut that separates signal (right) from background (left).}
\label{fig:tmva}       
\end{figure}

The TMVA package \cite{tmva} has been employed to try different classifiers. One that suited well ALPHA's experimental needs is the \textit{Boosted Decision Tree}, or BDT. This classifier has been employed to analyze the hold phase only, since the duration of the ramp down phase is only of 10-15 seconds. After proper training on independent datasets, constituted solely of either cosmic rays or $\h$ annihilation events, the result of its application to test samples is shown in Fig.~\ref{fig:tmva}. The fraction of accepted signal events is $\varepsilon_{\text{Hold}} = 42\%$ of the total size of the test sample, while the background rate measured is $R = 4.5\,$mHz: given the aforementioned trigger rate, the background suppression is $2000$-fold.

\section{Analysis of Trapped Antihydrogen}
\begin{figure}[!h]
\includegraphics[scale=0.14]{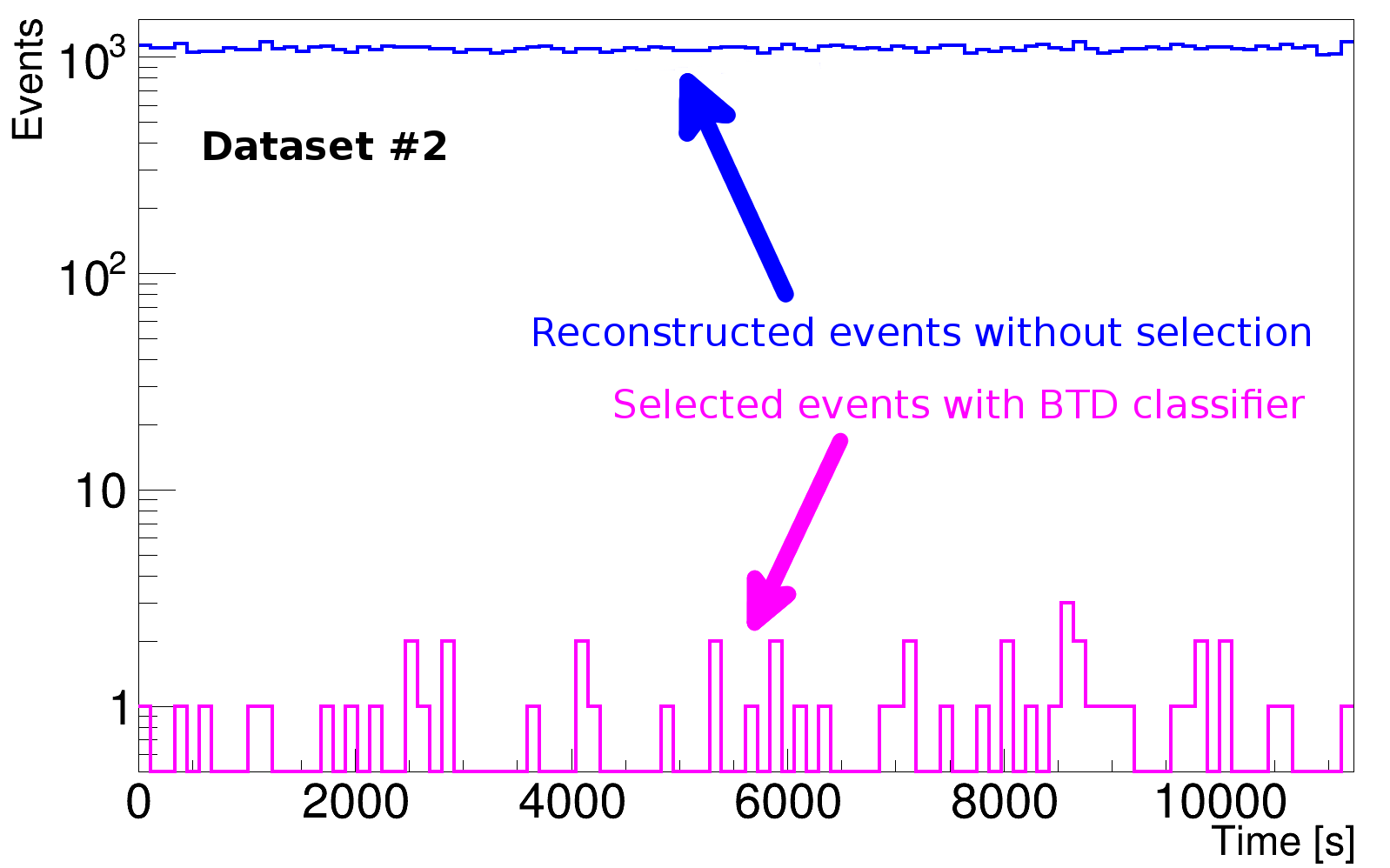}
\caption{Number of reconstructed events as a function of time (blue). Events that pass ML selection (magenta) are compatible with background only.}
\label{fig:HbarTrap}       
\end{figure}

The analysis the trapped $\h$ lifetime is conducted independently for each dataset, listed in Table~\ref{tab:exppar}. The number of expected background events in the hold phase, given in the third row of Table~\ref{tab:exppar}, is compatible, within the experimental uncertainty, with the number of events selected by the ML algorithm, displayed in the fourth row of the same table. The success of this sanity check permits the creation of histograms for each individual dataset, e.g., Fig.~\ref{fig:HbarTrap} for dataset \#2.

The aforementioned histograms are completed by including in the last bin the $\h$ counts detected during the ramp down phase, in order to exploit this datum in the likelihood fit. Each histogram is fitted with the piece-wise function, given by 
\begin{equation}\label{eq:fitfunc}
N = 
   \begin{cases}
    N_0 \, \lambda\Delta t \, \varepsilon_{\text{Hold}} \, \exp(-\lambda t) + B_{\text{Hold}} & \text{if } t < T\\
    N_0 \, \varepsilon_{\text{FRD}} \, \exp(-\lambda T) + B_{\text{FRD}} & \text{if } t = T  
   \end{cases}
\end{equation}
where the parameters to estimate are $N_0$, the number of trapped $\h$ and $\lambda$, the loss rate, while the common parameters to all four datesets are listed in Table~\ref{tab:anapar}. The duration of the hold phase $T$ for each dataset is shown in the second row of Table~\ref{tab:exppar}, while the number of expected background events in the ramp down phase $B_{\text{FRD}}$ is given in the last row.
\begin{table}[!h]
\caption{Fixed parameters used in Eq.~(\ref{eq:fitfunc})}
\label{tab:anapar}       
\resizebox{0.5\textwidth}{!}{  
\begin{tabular}{llc}
  \hline\noalign{\smallskip}
  \multicolumn{3}{c}{\textbf{Analysis Parameters}} \\
  \hline\noalign{\smallskip}
  $\varepsilon_{\text{Hold}}$     &Detection efficiency \textit{hold} phase              &42\%\\
  $\varepsilon_{\text{FRD}}$      &Detection efficiency \textit{ramp down} phase         &67\%\\
  $\Delta t$                      &Bin size in s                                         &10\\
  $B_{\text{Hold}}$               &Expected background (per bin) \textit{hold} phase     &$R\times\Delta t$ = 0.5\\
  \noalign{\smallskip}\hline
\end{tabular}
}
\end{table}

The results of the likelihood fit for each dataset is reported in Table~\ref{tab:fitres}.
The loss rate can be converted into a lifetime by considering the confidence interval bounds at, say, $1\sigma$ and taking the inverse values. The lower limit to the lifetime of magnetically trapped $\h$ is $\gtrsim 66\,$ hours (with the upper limit being infinite), while the total number number of trapped antiatoms is 1370.
 
It is worth noting that the total number of $\h$ released in the ramp down phase computed from the fifth row of Table~\ref{tab:exppar} and the detection efficiency $\varepsilon_{\text{FRD}}$ is compatible with $N_0$.

\begin{table}[!h]
\caption{Results of the likelihood fit of Eq.~(\ref{eq:fitfunc}) to the four datasets}
\label{tab:fitres}
\resizebox{0.45\textwidth}{!}{  
\begin{tabular}{lllll}
\hline\noalign{\smallskip}
                                                            &\textbf{\#1} &\textbf{\#2} &\textbf{\#3} &\textbf{\#4}\\
\noalign{\smallskip}\hline\noalign{\smallskip}
$\lambda\,[10^{-6}\,\text{s}^{-1}]$                         &1            &2            &8            &4\\
1$\sigma$ error on $\lambda\,[10^{-6}\,\text{s}^{-1}]$      &4            &6            &10           &8\\
\hline\noalign{\smallskip}
\textbf{Average loss rate}                                  &\multicolumn{4}{c}{$(2 \pm 2) \times 10^{-6} \; \text{s}^{-1}$}\\
\noalign{\smallskip}\hline\noalign{\smallskip}
\noalign{\smallskip}\hline\noalign{\smallskip}
$N_0$                                                       &547          &269          &215          &338\\
1$\sigma$ error on $N_0$                                    &32           &26           &18           &25\\
\hline\noalign{\smallskip}
\textbf{Total antiatoms trapped}                            &\multicolumn{4}{c}{$1370 \pm 51$}\\
\noalign{\smallskip}\hline
\end{tabular}
}
\end{table}

\section{Conclusions}

In this report a measurement of the lifetime of trapped $\h$ in the ALPHA apparatus has been presented. The rate at which $\h$ escapes the magnetic confinement is determined by comparing the number of $\h$ annihilation detected while the magnetic trap is being shut down to the number of events detected while the antiatoms are being confined. The latter is only possible by using advanced analysis technique, i.e., Machine Learning classifiers, in order to enhance the suppression of background events due to cosmic rays. The present measurement yields a lower limit of the lifetime of magnetically trap $\h$ of $2\times10^{5}\,$s or 66 hours. This preliminary result represents $200$-fold improvement with respect to the earliest ALPHA trapping campaigns \cite{AndreNat2011}, where seven $\h$ were held for a little over sixteen minutes.

\end{document}